# Generative Adversarial Networks for Brain Images Synthesis: A Review

Firoozeh Shomal Zadeh[1], Sevda Molani[2*], Maysam Orouskhani[1], Marziyeh Rezaei[3], Mehrzad Shafiei[1], Hossein Abbasi[4]

[1]Postdoctoral Research Fellow, Department of Radiology, University of Washington, Seattle, USA
Email: {Shomal, Maysam, Mshafie}@uw.edu

[2]Postdoc Researcher, Institute of Systems Biology, Seattle, USA (Corresponding Author)
Email: Smolani@systemsbiology.org

[3]Research Assistant, Department of Electrical and Computer Engineering, University of Washington, USA
Email: Marziyeh@uw.edu

[4] South Tehran Branch, Islamic Azad University, Iran
Email: Hossein.Abbasi48@gmail.com

**Abstract**: In medical imaging, image synthesis is the estimation process of one image (sequence, modality) from another image (sequence, modality). Since images with different modalities provide diverse biomarkers and capture various features, multi-modality imaging is crucial in medicine. While multi-screening is expensive, costly, and time-consuming to report to radiologists, image synthesis methods are capable of artificially generating missing modalities. Deep learning models can automatically capture and extract the high dimensional features. Especially, generative adversarial network (GAN) is one of the most popular generative-based deep learning methods, uses convolutional networks as generators, and estimated images are discriminated as true or false based on a discriminator network. This review provides brain image synthesis via GANs. We summarized the recent developments of GANs for cross-modality brain image synthesis including CT to PET, CT to MRI, MRI to PET, and vice versa.

## 1. Introduction

Artificial intelligence (AI) has aroused widespread interest in medical imaging. Especially with rapid progress in deep learning (DL) and development of the various image processing models, AI has turned into one of the hot topics of radiology research. Currently, enormous Conventional Neural Networks (CNN) are applied to different research and clinical applications, such as image segmentation, lesion detection, diagnosing, classification, and even interpretation. Multiple imaging modalities are used by radiologists to provide complementary information and a comprehensive description of a disease. Each of these expensive, time-consuming imaging modalities comes with disadvantages. Computed tomography (CT) images have high radiation risk, and positron emission tomography (PET) scans even entail additional radiation exposure. The long scanning time of magnetic resonance imaging (MRI) causes motion artifacts and results in low-resolution imaging. Novel DL-based approaches are applied to address these limitations by generating missing imaging modalities from available modalities. Image synthesis allows us to improve the quality and resolution of the imaging examinations and provide the opportunity to have more information and details about the disease in a time- and cost-effective manner [1-5]. In medicine, while statistical methods analyze the brain data clinically [6], deep neural networks are end-to-end learning models that automatically extract the high number of features in brain images including CT, MRI, and PET and are capable of learning complex pattern in which provide the great performance. Deep learning algorithms have also been utilized in brain image analysis. The applications of deep learning models in neuroimages can be mentioned as brain tumor classification and segmentation [7] for measuring and visualizing the brain's anatomical structures, analyzing brain changes, and detecting the shape of lesions or tumors in the brain. However, in some cases images with multi modalities provide different features as decision makers and bring the diverse features to consider. For example, to early detection of Alzheimer's disease at pre-clinical stage, screening of MRI and PET should be conducted simultaneously to analyze the various biomarkers. MRI analyses the anatomical structures of the brain while PET measure brain metabolism and amyloid tracers. Therefore, to make a true decision about the disease, using multi-modality to analyze features derived from different screening would be useful [8]. However, the main problem of multi-modality imaging is the cost of extra scan, radiation dose and delay clinical workflow. As a result, image synthesis (cross-modality image estimation) methods have been proposed to overcome these limitations. Images synthesis is the process of artificial generation of one image modality from different modality. Image synthesis techniques provide fewer scans, less delay, and lower radiation.

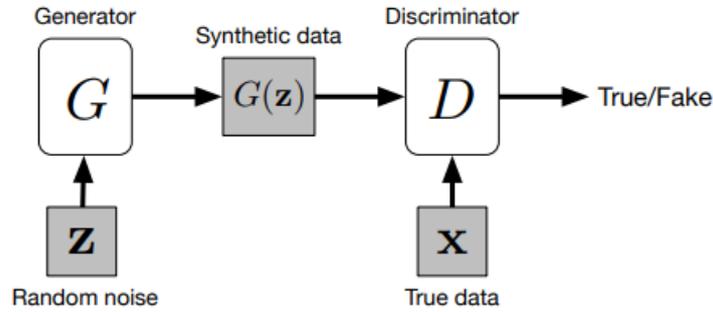

Fig. 1 The original GAN [22]

Recently, generative adversarial deep neural networks (GANs) [9] got a growing attention by the researchers. GAN is a class of machine learning frameworks including two neural networks: generative and discriminator part. While two networks contest with each other in a game, the generative network tries to produce the fake image so that the discriminator part cannot recognize the fake from the real image. GANs have been employed to tackle a wide array of challenges. One of the most prominent applications of GANs is image translation [10]. The main goal is to translate images between different techniques, as they can translate images across different modalities or generate new images within the same modality, but different sequences. For example, generating T1 sequence from T2. Since GANs can generate new data, they can be used for augmentation of brain images when we suffer from the lack of data. Super resolution to generate high resolution images from low resolution ones is another interesting application of GANs [11]. While noisy images are too challenging to interpret by the radiologists, GANs are used to remove noise and generate clear images [12]. The automatic segmentation of tumors and lesions in brain is another interesting application of GANs [13]. Finally, GANs are practical models to obtain the highly accurate reconstruction of natural images from brain activity [14].

In this paper, we review the different applications of GANs in brain image synthesis. The background section provides the definition of image synthesis in medical imaging and explains the basic model and the new versions of generative adversarial network. Section 3 describes the usage of GANs in brain image synthesis from CT to PET, CT to MRI, MRI to PET, and vice versa.

## 2. Background

*2.1. Generative Adversarial Networks*

Adversarial networks in general, and GANs more specifically, are trained to play a minimax game between a generator network which tries to maximize a certain objective function in tandem with a discriminator network which tries to minimize that same objective function hence the 'Adversarial' denomination. In their most basic formulation, GANs are trained to optimize the following value function [9,15]

(1)

$$\min_G \max_D V(D, G) = \mathbb{E}_{\boldsymbol{x} \sim p_{\text{data}}(\boldsymbol{x})}[\log D(\boldsymbol{x})]$$
$$+ \mathbb{E}_{\boldsymbol{z} \sim p_{\boldsymbol{z}}(\boldsymbol{z})}[\log(1 - D(G(\boldsymbol{z})))].$$

Here, $G(z)$ is the *generator network* with parameters $\theta_G$. It is fed with a random variable $z \sim p_z$ sampled from a given prior distribution that $G$ tries to map to $x \sim p_{data}$. To achieve this, another network $D$ with parameters $\theta_D$ is trained to differentiate between real samples $x \sim p_{data}$ from a given dataset and fake samples $\hat{x} \sim p\theta_G(x|z)$ produced by the generator. In doing so, the generator is pushed to gradually produce more and more realistic samples with the goal of making the discriminator misclassify them as real.

In order to handle the issues of the convergence speed, vanishing gradients, and model collapse, some modifications such as Deep Convolutional GAN, Least Square GAN [16], Wasserstein GAN [17], and Style GAN [18] have been proposed. Although this paper concentrates on cross-modality image synthesis and reviews generation of a modality from another modality (inter-modality) in brain, GANs have been used as image estimation for intra-modality applications. For example, in [19] authors translated the T1 sequence to T2 for reconstruction of high resolution from low resolution.

Synthesizing diffusion map from T1 via GAN was conducted by [20]. Moreover, authors in [21] synthesized 7T MRI from 3T MRI.

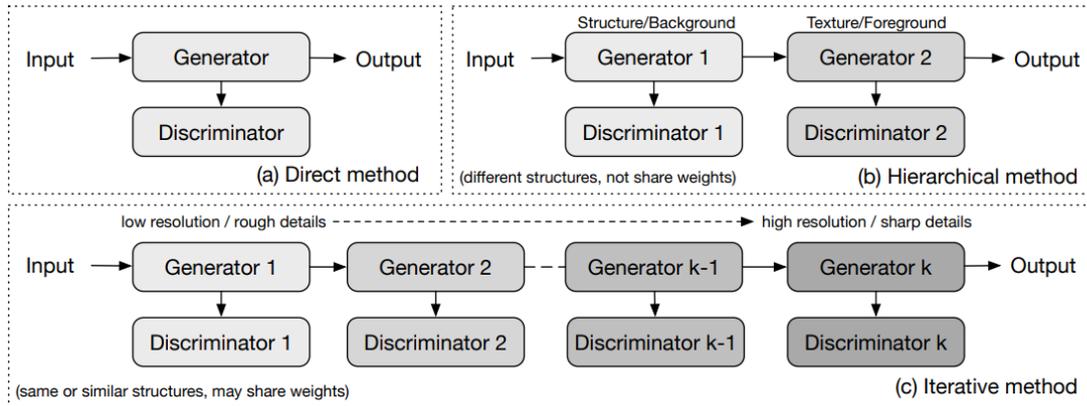

Fig. 2 Three approaches of image synthesis using Generative Adversarial Networks [22]

*2.2. Approaches of Image Synthesis with GAN*

GANs usually use three methods to generate fake images including Direct Methods, Iterative Methods and Hierarchical Methods [22]. The main difference comes from the number of generators and discriminators networks. While the Direct Method works with only one generator and one discriminator, the other two methods get benefit of using multiple generators and discriminators. In contrast to the Direct Method, algorithms under the Hierarchical Method such as SS-GAN [23] employ two networks for both generator and discriminator. These methods separate an image into two parts, like "styles & structure" and "foreground & background". The generators are connected through parallel or sequencing. However, the Iterative Methods use similar multiple generators, and they generate images from coarse to fine. In this model, generator (i) refines the results from the previous generator (i-1). Moreover, Iterative Methods take advantage of weight-sharing among the generators.

## 3. Brain Image Synthesis with GAN

In this section, we summarize different applications of GAN for brain image synthesis including MRI-CT, CT-PET, and MRI-PET as well.

*3.1. MRI-CT*

MR image synthesis from CT images is a challenging task due to large soft-tissue signal intensity variations. In this section, we review the methods which have been recently published to handle this issue. [5] proposed a Fully CNN combined with a cyclic GAN to generate CT images from MR images to reduce patients' radiation exposure. They conducted 315 images from the Alzheimer's Disease Neuroimaging Initiative (ADNI) dataset, and they successfully could generate CT images from MR images. MAE (Mean Absolute Error) and MSE (Mean Squared Error) were used to estimate the model consistency loss. After training the model, generator loss has come nearly equal to the discriminator loss. [26] developed a model to improve image synthesis performance using 3D Common-feature-learning-based Context-aware GAN (CoCa-GAN). They used encoder-decoder architecture to map available imaging modalities of glioma into a common feature space by the encoder to generate target missing imaging modalities by the decoder. Two different models: early-fusion-CoCa-GAN (eCoCa-GAN) and intermediate-fusion-CoCa-GAN (iCoCa-GAN), were compared to accommodate the common feature space; in their experiment, iCoCa-GAN outperformed eCoCa-GAN and finally recommended to develop the common feature space. They also conducted segmented images of the tumors to enhance synthesis tasks by emphasizing tumor regions. As segmentation tasks shared the same common feature space as synthesis tasks, given the input rough segmentation mask, allowed synthesis loss function to focus more on mass regions and represent the specific tumor information. This helped make the representation as similar as tumor appearance for image synthesis. Results indicated that iCoCa-GAN outperforms other models for image synthesis in terms of quality and improving the tumor segmentation, especially where available modalities are limited. [27] introduced a deep learning ResNet50-based CNN model to classify Alzheimer's disease using brain MRI. Besides, to increase the data set, they developed a CycleGAN model to generate MR images of the brain. They conducted a dataset of 705 samples labeled as normal cognition (NC) and 476 samples labeled as Alzheimer's disease (AD). The CycleGAn model with two generators synthesized NC samples from AD images and AD samples from the NC real images. Then the discriminator compared the fake synthetic images with the real ones to measure the GAN loss function. They successfully improved the accuracy of the Alzheimer's disease classification with promising progress in data synthesis.

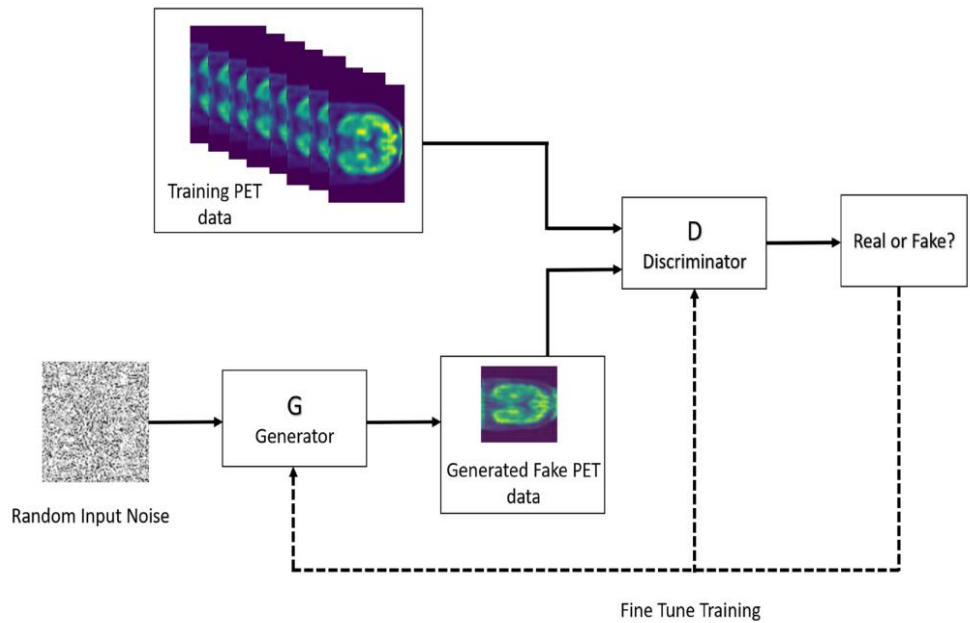

Fig. 3 An example of synthetic brain PET image generator [24]

[28] developed a switchable CycleGAN to augment cross-contrast MRI images and compare it to the original CycleGAN. Original CycleGAN needed 2 separate image generators (forward and backward generators) for the training phase, which required more time and parameters, while switchable CycleGAN used a switchable generator to synthesize images with different styles. Available Adolescent Brain Cognitive Development (ABCD), a large dataset of available brain MR images, was used to collect 1,517 subjects of T1- and T2-weighted images. Ten slices of each image were obtained, resulting in a total of 30,340 slices; of which 70% were used for training ,10% for testing, and 20% for testing. Peak signal-to-noise ratio (PSNR) and the structural similarity index measure (SSIM) were used for quantitative comparison of the 2 models and demonstrated that switchable CycleGAN outperformed the original Cycle GAN. In qualitative evaluation, which compared the visualization results of each model, switchable CycleGAN generated more consistent results with the target images with less artifacts and more details of brain tissue. Switchable CycleGAN also was found to be more robust on small datasets with less training time than the original CycleGAN. [29] generated a novel end-to-end hierarchical GAN architecture to augment high-resolution 3D images by conducting 9,276 thoracic CT and 3,538 brain MR images. Most AI model training is done by low-resolution images as the Graphical Processing Units (GPUs) memory is limited, which results in low-quality images with artifacts. The hierarchical structure is represented as a memory-efficient model, which simultaneously generates a low-resolution version of the images and a randomly selected sub-volume of the high-resolution images. The incorporated encoder enabled clinical-relevant feature extraction from sub-volume high-resolution images to ensure anatomical consistency and generate high-resolution images with reduced required memory for training. The performance of the model was explored both qualitatively and quantitatively. If fake images were similar to real images, the quality of the image was quantitatively assessed by Frechet Inception Distance (FID), Maximum Mean Discrepancy (MMD), and Inception Score (IS). The hierarchical GAN model achieved better results in qualitative and quantitative analysis and could generate more realistic images than the baseline model. [30] adopted pix2pix with a 3D framework of the GAN model to augment CT images from contrast-enhanced MR images. Their study aimed to generate CT images to help plan radiotherapy treatment. The model is also designed to improve the quality and resolution of the generated CT images. They used 26 paired CT and MRI scans to train their model and, the rest 5 paired were used as a testing set. The generated scan's similarity to real images was evaluated by quantized image similarity formulas, including cosine angle distance, Euclidean distance, mean square error, PSNR, and SSIM. The satisfaction rated by radiologists was excellent for spatial geometry and noise level, good for contrast and artifacts, and fair for anatomical and structural details. [31] developed a GAN-based model to generate MRI from CT scan to detect acute ischemic stroke in suspected patients. They hypothesized that the diagnostic accuracy of brain lesions would increase by using synthetic MRI instead of CT. They used 140 examinations for training and 53 imaging examinations for testing to build the pix2pix GAN framework. In their model, the generator used CT images to synthesize MRI while the discriminator took synthetic or real MRI to assess whether it was fake or real. A neuroradiologist with 9 years of practical experience assessed the quality of the synthetic MRI visually, and no significant structural or signal intensity differences were found. PSNR and SSIM were used to estimate the similarity of the synthetic images to real onesRegarding reader performance in patient selection, the sensitivity increased by using Synthetic MRI than CT, but the specificity decreased. They concluded that the GAN model has the potential to generate MR images from non-contrast CT scans and can improve the sensitivity of acute stroke detection. Although, the image similarity performance was poor and, further expert discrimination was recommended to enhance the correctness of synthetic images.

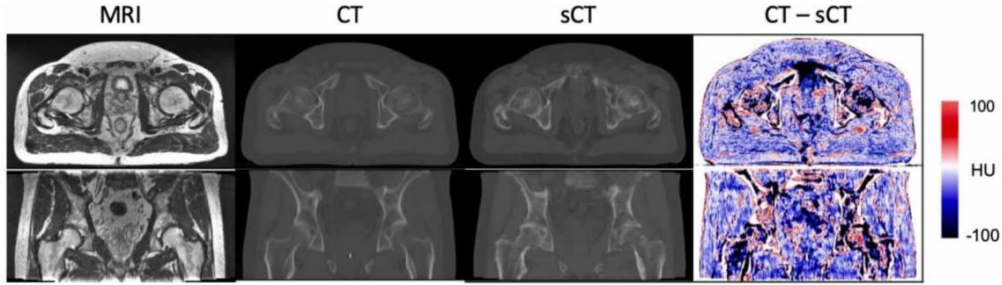

Fig. 4 An example of sCT estimated from MRI [25]

Table 1. A Summary of MRI-CT Image Synthesis

| Author (year) | Method | Input | Estimation | Architecture | Dataset | Application | Metric |
|---|---|---|---|---|---|---|---|
| **Huang (2022)** | eCoCa- and iCoCa-GAN | Synthetic MRI | Target missing | 3D CoCa-GAN | 335 MRI of MICCAI-BraTS 2019 | Diagnosis and treatment of glioma | PSNR, NMSE, SSIM |
| **Badr (2021)** | ResNet50-based CNN | MRI | MRI | CycleGAN | 1,181 MRI from IITP date set | Alzheimer's classification | Accuracy |
| **Zhang (2022)** | switchable and original CycleGAN | T2WI | T1WI | Switchable CycleGAN | 1,517 MRI from ABCD data set | Enhance image synthesis quality | PSNR and SSIM |
| | | T1WI | T2WI | | | | |
| **Nehra (2021)** | Cycle GAN | MRI | CT | FCN and GAN | 315 MRI from ADNI date set | N/A | MAE, MSE |
| **Sun (2021)** | Hierarchical Amortized GAN | MRI | MRI | End-to-end GAN | 3,538 MRI from GSP date set | High-resolution Images | FID, MMD, IS |
| **Wang et al. (2022)** | pix2pix with a 3D framework | MRI | CT | GAN | 31 paired CT and MRI of Chang Gung Memorial Hospital | Radiotherapy planning | CAD and MSSIM |
| **Na Hu et al. (2022)** | 3D-CT2MR | CT | MRI | GAN | 193 | Detection ischemic stroke | PSNR, SSIM |

\* eCaCa- and iCoCa-GAN: early-fusion intermediate-fusion Common-feature learning-based Context-aware, PSNR: peak signal-to-noise ratio, SSIM: the structural similarity index measure, IITP: Institute for Information & Communications Technology Promotion, ABCD: Adolescent Brain Cognitive Development, GSP: Brain Genomics Superstruct Project, FID: Frechet Inception Distance, MMD: Maximum Mean Discrepancy, IS: Inception Score, CAD: cosine angle distance, MSSIM: mean structural similarity index

### 3.2. MRI-PET

Magnetic Resonance Image (MRI) and Positron emission tomography (PET) are used to diagnose a wide range of diseases. PET imaging is expensive and not offered in most of the medical centers in the world due to its high cost and increased risk of radiation exposure. PET synthesis from MRI multi-modal images has become a popular method that can reduce the cost and patient's radiant dose caused by PET imaging. [32] proposed a 3D self-attention conditional GAN named SC-GAN by extending a 2D conditional GAN into a 3D conditional GAN and adding a 3D self-attention module to it to generate PET synthetic images from MRI scans. A self-attention module models the relationship between widely separated image voxels which helps to improve the quality of generated images and reduce the blurriness. Exhaustive loss functions were used in this method like spectral normalization, feature matching loss, and brain area RMS error (RMSE) that improved the accuracy of image synthesis. The dataset used in this work was obtained from Alzheimer's Disease Neuroimaging Initiative 3 (ADNI-3). The input was MRI scans selected from T1-weighted (T1w) and fluid-attenuated inversion-recovery (FLAIR) structures and the target were PET scans selected from amyloid PET. 265 subjects were selected where 207 of them were used for training and 58 of them were used for testing. The model was then evaluated by comparing NRMSE, PSNR, and SSIM metrics with other works.

Table 2. A Summary of MRI-PET Image Synthesis

| Author (year) | Method | Input | Estimation | Architecture | Dataset | Application | Metric |
|---|---|---|---|---|---|---|---|
| **Lan (2021)** | SC-GAN | MRI | PET | Conditional GAN | ADNI | Multimodal 3D Neuroimaging Synthesis | NRMSE PSNR SSIM |
| **Sikka (2021)** | GLA-GAN | MRI | PET | GAN | ADNI | Diagnosis of Alzheimer's Disease | MAE PSNR SSIM |
| **Shin (2020)** | GANBERT | MRI | PET | GAN | ADNI | MRI to PET Image Synthesis | PSNR SSIM RSME |
| **Wei (2019)** | Sketcher-Refiner GAN | MR | PET | Conditional GAN | Clinical Dataset | Myelin Content Prediction | MSE PSNR |
| **Pan (2018)** | 3D Cycle-Consistent GAN | MRI | PET | CycleGAN | ADNI | Diagnosis of Alzheimer's Disease | PSNR |
| **Zhang (2022)** | BPGAN | MRI | PET | U-Net | ADNI | Multi-Modal Medical Imaging | MAE PSNR SSIM |

[33] focused on the cross-modality synthesis of PET scans from MRI images using globally and locally aware image-to-image translation GAN (GLA-GAN) with a multi-path architecture for Alzheimer's disease (AD) diagnosis. It was assumed that by exploiting both global and local contexts, the quality of synthesized PET scans can be improved. In this work, SSIM (MS-SSIM) was used as an additional objective function to improve synthetic image quality. 402 input and target samples were selected from the ADNI dataset having both prepossessed MRI and FDG-PET modalities for training procedure. Finally, the quality of the synthesized images and the model accuracy for AD diagnosis were evaluated using SSIM, PSNR, and MAE metrics to compare with other works. [34] proposed a new method named GANBERT to generate PET images from MRI scans in a wide intensity range. The architecture is composed of a 3D U-Net-like generator that generates PET images from the MRI scans. It also has two Bidirectional Encoder Representations from Transformers (BERT) that are trained to predict real and synthetic PET images where its next sentence prediction (NSP) acts as a GAN discriminator. ADNI dataset was used to train and test the proposed model. Target was selected from 2,387 Amyloid PET (AV45), 536 Tau PET (AV1451), and 3,108 fluorodeoxyglucose PET (FDG) that were paired with T1- weighted input MRI images. Then the model was evaluated by comparing the quality of generated PET images with other methods using PSNR, SSIM, and RMSE metrics. [35] introduced a new approach called Sketcher-Refiner GAN consisting of two conditional GANs to predict the PET-derived myelin content from multimodal MR images. The Sketcher network generates global anatomical information, and the Refiner network calculates the tissue myelin content. The dataset used in this work was MRI and PET images collected from 18 MS patients and 10 age-matched healthy volunteers. The generated image quality was compared with the state-of-the-art methods using MSE and PSNR. They also compared myelin content prediction in three: white matter (WM) in healthy controls (HC), normal appearing white matter (NAWM) in MS patients, and lesions in MS patients ROIs with other works showing noticeable improvement on prediction accuracy. [36] proposed a new method called 3D Cycle-consistent GAN which is a two-stage deep learning method for AD diagnosis using MRI and PET data. First, PET images were generated from the corresponding MRI data by using 3D Cycle-consistent Generative Adversarial Networks (3D-cGAN). Then a deep multi-instance neural network was implemented for AD diagnosis and mild cognitive impairment (MCI) prediction using the synthetic PET and MRI images. The proposed model was evaluated using two ADNI sub-datasets, ADNI-1 and ADNI-2. The model was first trained by ADNI-1 (containing both PET and MRI) images and then tested on the complete subjects in ADNI-2. The quality of the synthetic images was then evaluated using the PSNR metric and the experimental results demonstrated that the synthetic PET images produced by this method were reasonable. [37] proposed BPGAN which is, a 3D end-to-end generative adversarial network, that can synthesize brain PET images from MRI scans for multi-modal medical imaging research. They designed a 3D multiple convolution U-Net (MCU) generator to improve the quality of synthesized PET images and then employed a 3D gradient profile (GP) and structural similarity index measure (SSIM) loss functions to gain higher similarity to the ground truth images. They tested their model on ADNI database and evaluated it by using mean absolute error (MAE), PSNR and SSIM metrics. Qualitative evaluations demonstrated improvement on multi-class AD diagnosis accuracies compared to the stand-alone MRI.

Table 3. A Summary of CT-PET Image Synthesis

| Paper | Method | Input | Estimation | Architecture | Dataset | Application | Metric |
|---|---|---|---|---|---|---|---|
| Armanious K, et.al 2020) | MedGAN | PET | CT | cGAN | SOMATOM mCT, Siemens Healthineers, Germany | Image-to-image translation | SSIM, PSNR (dB), MSE, VIF, UQI, LPIPS, |
| Armanious K, et.al (2020) | MedGAN | PET | CT | cGAN | 90 patients with Fluorine-18-FDG PET/CT scans of the head region | Attenuation correction | Mean difference |
| Rao et. Al (2022) | Image synthesis network and nonrigid registration | PET | pseudo-CT | GAN | 25 patients for training. 12 patients for evaluation | Attenuation correction | PSNR, MAPE |
| Liu et. Al (2018) | deepAC | PET | pseudo-CT | data-driven deep learning approach | Discovery PET/CT 710 scanner (GE Healthcare, Waukesha, WI, USA) | Attenuation correction | Dice coefficient, MAE |

*3.3. CT-PET*

Synthesizing CT from PET images is challenging due to less resolution and detailed information in PET images compared to CT images [38]. Despite these challenges, several studies were able to generate results with low average errors. [39] A new proposed GAN framework (MedGAN) combines the fragmented benefits of several translation approaches such as ResNets, pix2pix, PAN and Fila-sGAN with a new high-capacity generator architecture. The purpose of this framework is to improve technical post-processing tasks that require globally consistent image properties with an application in PET-CT translation. Furthermore, they incorporated non-adversarial losses such as the perceptual, style and content losses as part of the framework. To test this framework, a dataset of 46 patients of the brain region acquired on a joint PET/CT scanner was used. The proposed framework produced realistic and homogeneous structures in the CT images that closely matched the ground truth CT images. [40] In another study adapts this proposed framework and evaluate the MedGAN framework for independent attenuation correction of brain fluorine-18-fluorodeoxyglucose (F-FDG) PET images only based on non-attenuation corrected PET data (NAC PET). In this study, a dataset consisting of NAC PET and the corresponding CT data from 50 patients were used for training and the information from 40 patients were used for technical and clinical validation. The results show that independent attenuation correction of brain F-FDG PET is feasible with high accuracy using the proposed framework. [41-42] proposed a framework to obtain attenuation information in a delayed clinical PET scanner without the need for additional CT scans. For this purpose, a GAN-based image synthesis network is developed to convert the PET back projection (BP) image and the NAC PET image into a pseudo-CT image. Later a non-rigid registration is performed between the CT image of the first scan and this pseudo-CT image to obtain the transformation field between the two scans. The final estimated CT image for the delayed PET image is obtained by applying the transformation field onto the CT images of the first scan. In this study, experiments with clinical datasets are implemented to assess the effectiveness of the proposed method with the Generative Adversarial Networks (GAN) method.

## Conclusion

In this paper, we reviewed the generative adversarial network and its applications to brain image synthesis. We also summarized the methods of generating artificial images used by GANs including Direct Methods, Hierarchical Methods, and Iterative Methods. Since GANs are composed of two different networks, generative and discriminative network, are powerful deep learning-based models to generate images and synthesize medical images. We then categorized the applications of GANs for brain image synthesis into three classes: CT-MRI, MRI-PET, and CT-PET and reviewed each method separately.